\documentclass[11pt]{article}
\usepackage{amsmath}
\usepackage{cite}
\usepackage{float}
\usepackage[T1]{fontenc}
\usepackage{subfigure}
\usepackage{graphicx}
\usepackage{epsfig}

\begin{document}

\title{\textbf{Unifying thermodynamic and kinetic descriptions of single-molecule processes: RNA unfolding under tension}}
\author{J.M. Rubi,$^a$ D. Bedeaux$^b$ and S. Kjelstrup$^{b}$ \\
$^a$ Departament de Fisica Fonamental, Universitat de Barcelona,\\
Diagonal 647, 08028 Barcelona, Spain\\
$^b$Department of Chemistry, Faculty of Natural Science and Technology,\\
Norwegian University of Science and Technology, Trondheim, 7491-Norway}
\maketitle
\date{}

\begin{abstract}
\noindent We use mesoscopic non-equilibrium thermodynamics theory to describe RNA unfolding under tension. The theory introduces reaction coordinates, characterizing a continuum of states for each bond in the molecule. The unfolding considered is so slow that one can assume local equilibrium in the space of the reaction coordinates. In the quasi-stationary limit of high sequential barriers, our theory yields the master equation of a recently proposed sequential-step model. Non-linear switching kinetics is found between open and closed states. Our theory unifies the thermodynamic and kinetic descriptions and offers a systematic procedure to characterize the dynamics of the unfolding process. 


\end{abstract}

\section{Introduction}

The understanding of how biomolecules fold to reach their functional state, is a fundamental problem.  Folding is the result of competition between innumerable intermolecular events, to eventually reach the state of lowest energy.  A particularly interesting case is the folding of ribonucleic acids (RNA), due to their vital role in the cell for information transfer, regulation, and catalysis. The RNA molecule is relatively simple, and the folding process is hierarchical \cite{Tinoco.Bustamante}. Experiments using a force to unzip the RNA helix-loop, have shown that each base pair in the molecule undergoes a transition between a zipped and an unzipped state, see Fig.1. This transition has been modelled by the kinetics of crossing large energy barriers which separate the closed and open conformations \cite{Cocco}, \cite{Vieregg.Tinoco}. 

Biomolecular processes have been approached using classical thermodynamics and reaction kinetics \cite{Tinoco}. Classical thermodynamics gives equilibrium quantities such as energies of states. Classical kinetic studies give transport properties such as activation energies and rate constants. A minimal kinetic model, which can account for many processes taking place in biomolecules, is the activated jump model for a single energy barrier, see e.g. \cite{looping.current}, \cite{looping.pnas}, \cite{Hill2}, \cite{Qian}. 

Both classical studies are useful, but one may ask if there is a more general basis for both studies? The answer is not obvious as classical non-equilibrium thermodynamics, a kinetic theory with a thermodynamic basis, deals with linear processes only \cite{Caplan}, \cite{Westerhoff}, while kinetic results indicate that highly non-linear phenomena are involved. Furthermore, how can one apply a thermodynamic analysis to the events in \textit{one} molecule?  

There are arguments in favour of a positive answer to these questions, however. Firstly, it was shown by Hill \cite{Hill}, that despite the lack of a thermodynamic limit, small systems such as biomolecules can be dealt with, using the principles of equilibrium thermodynamics \cite{Callen}. Thermodynamics of small systems was thus used to describe stretching experiments performed with single-molecules \cite{stretching}. Secondly, many experiments are performed with the molecule embedded in a heat bath. For this reason it undergoes Gaussian fluctuations, which makes the underlying stochastic process compatible with non-equilibrium thermodynamics principles \cite{Onsager.Machlup}. 

\begin{figure}
	\centering
		\includegraphics[height=3in]{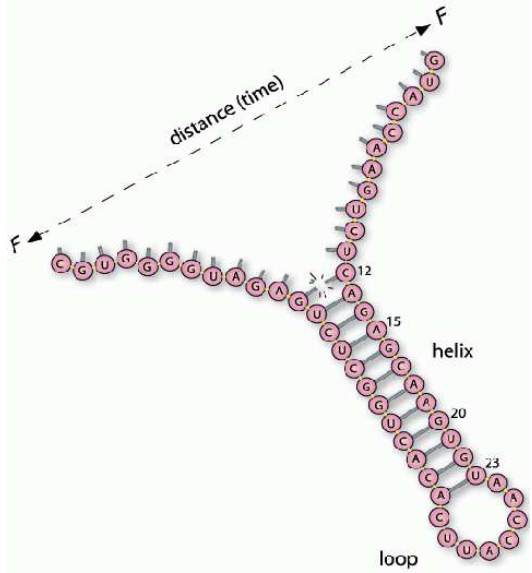}
	\caption{The unfolding of RNA by means of a force $F$. The numbering of base pairs start at the open end. Bond no 12 is being broken.}
	\label{fig:RNA-Fig01-9-colour.eps}
\end{figure}

It has thirdly been shown that when non-equilibrium thermodynamics is applied on the meso-scale, i.e. to small distances and short time domains, one can perfectly well account for the full non-linear dynamics of the process. The theory proposed, called mesoscopic non-equilibrium thermodynamics, has been applied to different situations involving activated processes \cite{mnet}, \cite{JTB}, \cite{PCCP}. A  simple feature explains this possibility: by adding the linear contributions to the rate, obtained when the system proceeds along the path connecting the initial and the final state, a non-linear behavior is found. The restriction of giving linear relationships does therefore not invalidate the use of non-equilibrium thermodynamics methods for the description of activated processes. 

In this paper, we will thus show that when such a mesoscopic description is applied to RNA unfolding, one can reproduce the master equation proposed in a kinetic description \cite{Cocco}, \cite{Vieregg.Tinoco}. 

The paper is organized in the following way. In section 2, we present the sequential-step model used to analyze the unzipping kinetics of a RNA molecule subject to an external force. In Section 3, we show that each step in this process can be viewed as a one-dimensional diffusion process in which the diffusion current is the unfolding rate. The linear regime of this quantity is discussed in Section 4 showing that classical non-equilibrium thermodynamics provides a description of this regime. Our approach to the kinetics of the process is presented in Section 5. Finally, in Section 6 we summarize the main characteristics of the theory presented.

\section{Activation over sequenced barriers}

The sequential-step model \cite{Cocco}, \cite{Vieregg.Tinoco} for RNA folding under tension (forces above a few pN) describes the opening and closing of base pairs at the boundary between the doubled-stranded region and the single-stranded end to which the force is applied. 
The molecule can then be found in different states identified by the terminal closed base pair $n$, where state $n=1$ is the completely folded state and state $n=N$ is the completely unfolded state. The total number of base pairs is $N-1$. Fig. 1 shows the configuration of the molecule in one of these states.

The model assumes that the process can be described by the transiton-state theory with the opening and closing rates of the base pair n upon application of an external force $F$, respectively given by
\begin{equation}
k_{o}(n)= \frac{1}{\tau}\exp({-\frac{g_{o}(n)}{k_{B}T}}) ;  \text{\ \ \ } k_{c}(n)=\frac{1}{\tau} \exp({-\frac{g_{c}(n,F)}{k_{B}T}})
\end{equation}
Here $g_{o}(n)$ is the energy necessary to open base pair $n$ at zero force, $g_{s}(n,F)$ is the energy associated with stretching the newly single-stranded section of the molecule by the force, and $\tau$ is a microscopic time corresponding to the diffusion time of a few-nm-size objects \cite{Doi.Edwards}. The model assumes that the energy needed to bring two nucleotides close enough to bond is large in comparison with $k_{B}T$.

The dynamics of the model was described in terms of the probability $\rho_{n}(t)$ for the molecule to be in state $n$ at time $t$. The evolution of the probability in time followed the master equation \cite{Cocco}
\begin{equation}
\dot{\rho}_{n}=-[k_{o}(n+1)+k_{c}(n,F)]\rho_{n}+k_{o}(n)\rho_{n-1}+k_{c}(n+1,F)\rho_{n+1}
\end{equation}
with the condition $\rho_{0}=\rho_{N+1}=0$. 
This equation is similar to the one used to describe nucleation kinetics, where the rates correspond to adsorption and desorption of particles to a nucleating cluster \cite{master.nucleation}.

\section{The unfolding rate}

A description of the dynamics of the folding process can be found through the conservation law
\begin{equation}
\dot{\rho}_{n}=-j_{n}+j_{n-1}
\end{equation}
where the unfolding rate of the nth base pair is given by
\begin{equation}
j_{n}=-k_{c}(n+1,F)\rho_{n+1}+ k_{o}(n+1)\rho_{n}  
\end{equation}
and $j_{0}=0$ for $n=1$. The conservation law toghether with the expression for the folding rate coincides with the master equation (2). 

An interpretation of the unfolding rate can be given from its expression in the continuum limit in which $\rho_{n}(t)\rightarrow\rho(n,t)$ and $\rho_{n}(t)-\rho_{n-1}(t)\rightarrow\partial\rho(n,t)/\partial n$. Writting $j_{n}$ in the equivalent form 
\begin{equation}
j_{n}=-\left[k_{c}(n+1,F)-k_{o}(n+1)\right]\rho_{n+1}- k_{o}(n+1)\left(\rho_{n+1}-\rho_{n}\right)
\end{equation}
we conclude that the term inside the square bracket can be interpreted as minus a velocity, i.e. a net progression from one state to the next, and $k_{o}(n+1)$ as a diffusion coefficient for the time rate of change of the probability density of state $n$. 

In the stationary state, the unfolding rate vanishes and when Eqs. (1) and (4) are used it implies that the condition of detailed balance 
\begin{equation}
\frac{\rho_{n+1,st}}{\rho_{n,st}}=\exp({-\Delta G(n,F)/k_{B}T})
\end{equation}
is valid. Here $\Delta G(n,F)=g_{o}(n)-g_{s}(n,F)$ is the energy change upon opening the nth base pair. This condition ensures that the stationary probability density of state $n$ of the molecule is given by
\begin{equation}
\rho_{n,st}=\rho_{1,st} \exp({-G(n,F)/k_{B}T})
\end{equation}
where $G(n,F)$ is the stationary value of the energy of state n relative to state 1, given $F$,
\begin{equation}
G(n,F)=\sum_{m=1}^{n-1}\Delta G(m,F)
\end{equation}
and $G(1,F)=0$.

\begin{figure}
	\centering
		\includegraphics[height=3in]{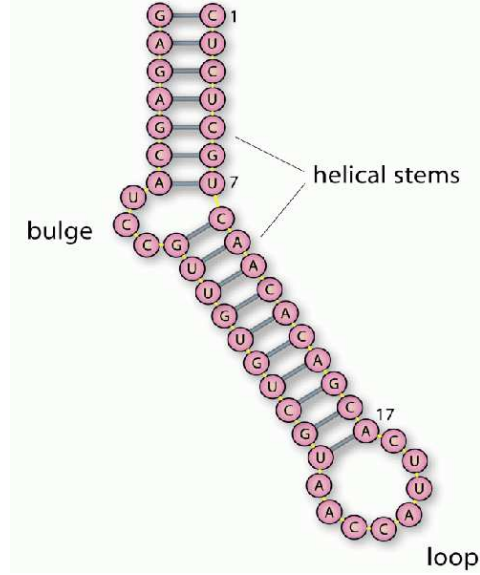}
	\caption{A RNA strand, with a bulge at $n$= 7 and altogether 17 pairs to break}
	\label{fig:RNA-Fig2}
\end{figure}

\begin{figure}
	\centering
		\includegraphics[height=5in]{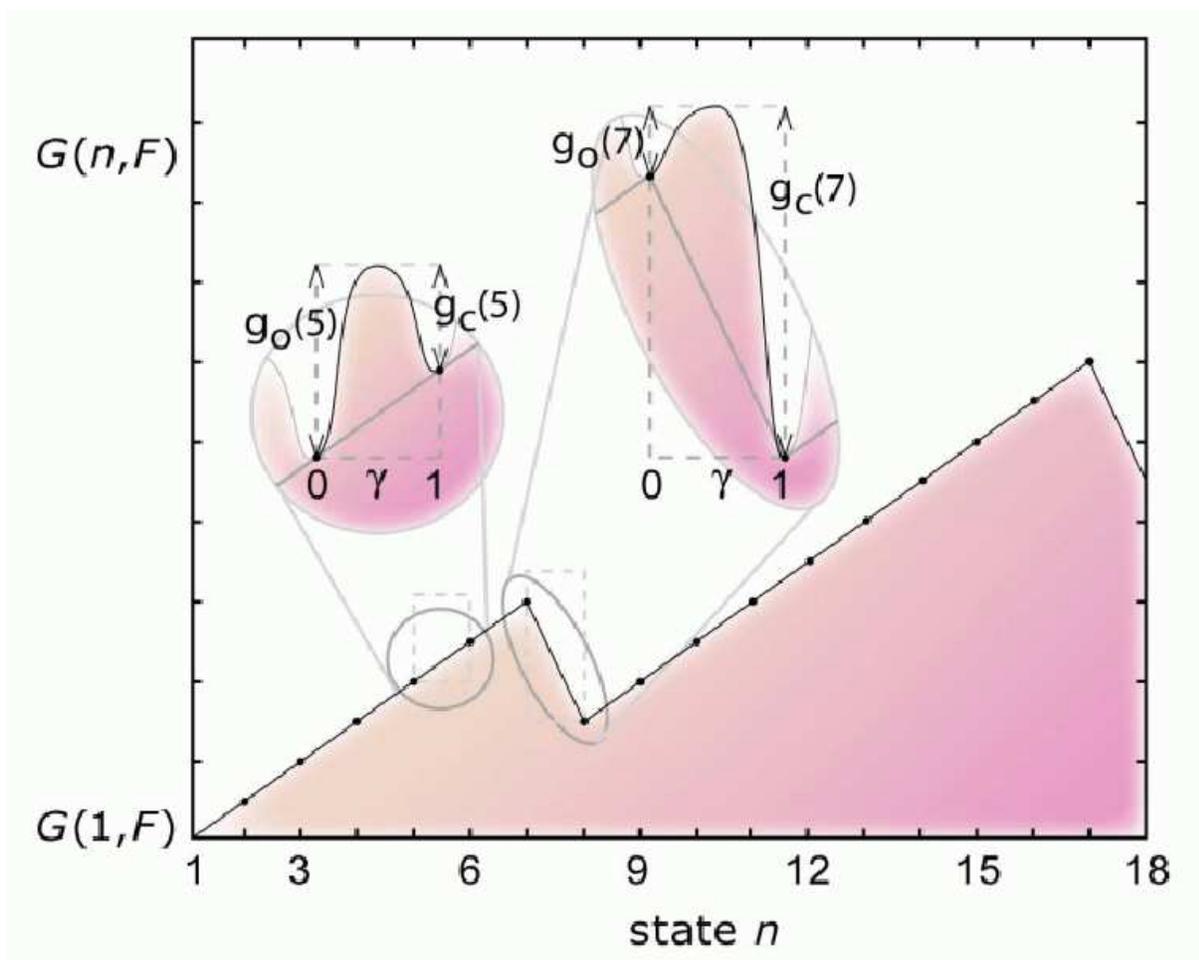}
	\caption{The molecular Gibbs energy as a function of the molecular state as given by the number of pairs broken ($n$) and the force used ($F$). Inserts show the reaction coordinate.}
	\label{fig:RNA-Fig3}
\end{figure}

Consider for the sake of illustration a more specific example than given in Fig.1, namely Fig.\ref{fig:RNA-Fig2}. The molecule has a bulge at $n$= 7, and there are altogether 17 pairs to break.
The molecular Gibbs energy of the molecule in  Fig.\ref{fig:RNA-Fig2} is pictured in Fig.\ref{fig:RNA-Fig3}. We see the continuous increase in energy as the pairs are breaking, and the decreases caused by the bulge and the loop. The inserts in  Fig.\ref{fig:RNA-Fig2} give more details on the path taken, as one bond is breaking. In particular we see the path across the activation energy barrier. The heigth to cross to open a bond is $g_o(n)$, while the heigth to cross to close a bond is $g_c(n)$.  

The chemical potential relative to the stationary state is
\begin{equation}
\mu_{n}=k_{B}T\ln\frac{\rho_{n}}{\rho_{n,st}}
\end{equation}
The unfolding rate can then be expressed by
\begin{equation}
j_{n}=-\frac{\rho_{n,st}}{\tau}\exp({-\frac{g_{o}(n)}{k_{B}T}})\left(\exp({\frac{\mu_{n+1}}{k_{B}T}})-\exp({\frac{\mu_{n}}{k_{B}T}})\right)
\end{equation}
which has the form of the law of mass action.

\section{Non-equilibrium thermodynamics for the linear regime of the folding rate}

We show in this section that classical non-equilibrium thermodynamics \cite{dGM} provides a description of the process only in the case when the state of the molecule is not far from the stationary state \cite{Hill2}. In this regime, the unfolding rate is given through the linear approximation to Eq. (10) and is therefore proportional to the difference between base pair chemical potentials. To obtain the explicit expression of the rate under this condition, we use the statistical entropy per molecule
\begin{equation}
S(t)=S_{st}-k_{B}\sum_{n=1}^{N}\rho_{n}\ln\frac{\rho_{n}}{\rho_{st}}
\end{equation}
which also can be expressed in terms of the chemical potentials as
\begin{equation}
S(t)=S_{st}-\frac{1}{T}\sum_{n=1}^{N}\rho_{n}\mu_{n}
\end{equation}
Here $S_{st}$ is the entropy of the molecule in the stationary state. By taking the time derivative of $S(t)$ and using the normalisation condition:
\begin{equation}
\sum_{n=1}^{N}\rho_{n}=1
\end{equation}
we obtain the entropy production rate 
\begin{equation}
\dot{S}(t)=-k_{B}\sum_{n=1}^{N}\dot{\rho}_{n}\ln\frac{\rho_{n}}{\rho_{n,st}}=-\frac{1}{T}\sum_{n=1}^{N}\dot{\rho}_{n}\mu_{n}
\end{equation}
Using now the probability conservation law (3), this quantity can be expressed as
\begin{equation}
\dot{S}(t)=-\frac{1}{T}\sum_{n=1}^{N}(-j_{n}+j_{n-1})\mu_{n}
\end{equation}
or likewise as
\begin{equation}
\dot{S}(t)=-\frac{1}{T}\sum_{n=1}^{N}j_{n}(\mu_{n+1}-\mu_{n})
\end{equation}
From the entropy production rate and in accordance with Onsager's theory we can infer the value of the current
\begin{equation}
j_{n}=-\frac{1}{T}\sum_{m}l_{nm}(\mu_{m+1}-\mu_{m})
\end{equation}
where the coefficients $l_{nm}$ are Onsager coefficients.
If  the unfolding rate is only coupled to its own conjugate thermodynamic force, we obtain
\begin{equation}
j_{n}=-\frac{l_{n}}{T}(\mu_{n+1}-\mu_{n})
\end{equation}
where we have defined $l_{n}\equiv l_{nn}$. If we identify the Onsager coefficient with 
\begin{equation}
l_{n}=\frac{\rho_{n,st}}{\tau k_{B}}\exp({-\frac{g_{o}(n)}{k_{B}T}})
\end{equation}
this expression coincides with the linear approximation to Eq.(10). In the next section, we show that the complete kinetics of the unfolding process follows from a non-equilibrium thermodynamics scheme formally similar to the one used here.

\section{Kinetic description of RNA unfolding based on thermodynamic grounds}

A complete description of the RNA unfolding process can be given if we observe transitions between states on shorter time scales, or, when the switching kinetics over each barrier can be viewed as a diffusion process. The slow evolution of the system over the barriers in the time scale considered, justifies the assumption that during the opening or closing of base pair n the molecule passes through a sequence of local equilibrium states characterized by the reaction coordinate $\gamma_{n}$. The coordinate was illutrated in the insert of Fig.3 for $n$=5 and 7.

To obtain the form of the $n$'th unfolding rate $j_{n}$, we follow the procedure indicated in Section 4. Our starting point is the statistical expression for the entropy 
\begin{equation}
S(t)=S_{st}-k_{B}\sum_{n=1}^{N}\int_{0}^{1} d\gamma_{n}\rho_{n}(\gamma_{n},t)\ln\frac{\rho_{n}(\gamma_{n},t)}{\rho_{n,st}(\gamma_{n})}
\end{equation}
which can also be written as
\begin{equation}
S(t)=S_{st}-\frac{1}{T}\sum_{n=1}^{N}\int_{0}^{1} d\gamma_{n}\rho_{n}(\gamma_{n},t)\mu_{n}(\gamma_{n},t)
\end{equation}
where the chemical potential along the $\gamma_{n}$ coordinate relative to the stationary state is given by
\begin{equation}
\mu_{n}(\gamma_{n},t)=k_{B}T\ln\frac{\rho_{n}(\gamma_{n},t)}{\rho_{n,st}(\gamma_{n})}
\end{equation}
The normalization condition in this description is:
\begin{equation}
\sum_{n=1}^{N}\int_{0}^{1}d\gamma_{n}\rho_{n}(\gamma_{n},t)=1
\end{equation}
To calculate the entropy production rate, we follow the procedure indicated in the previous section. From the time derivative of this expression, we obtain
\begin{equation}
\dot{S}(t)=-k_{B}\sum_{n=1}^{N}\int_{0}^{1} d\gamma_{n}\dot{\rho}_{n}(\gamma_{n},t)\ln\frac{\rho_{n}(\gamma_{n},t)}{\rho_{n,st}(\gamma_{n})}=-\frac{1}{T}\sum_{n=1}^{N}\int_{0}^{1} d\gamma_{n}\dot{\rho}_{n}(\gamma_{n},t)\mu_{n}(\gamma_{n},t)
\end{equation}
Using now the probability conservation law 
\begin{equation}
\dot{\rho}_{n}(\gamma_{n},t)=-\frac{\partial}{\partial\gamma_{n}}j_{n}(\gamma_{n},t)
\end{equation}
we get the equivalent expression
\begin{equation}
\dot{S}(t)=\frac{1}{T}\sum_{n=1}^{N}\int_{0}^{1} d\gamma_{n}\mu_{n}(\gamma_{n},t)\frac{\partial}{\partial\gamma_{n}}j_{n}(\gamma_{n},t)
\end{equation}
Partial integration then yields the entropy production rate
\begin{equation}
\dot{S}(t)=-\frac{1}{T}\sum_{n=1}^{N}\int_{0}^{1} d\gamma_{n}j_{n}(\gamma_{n},t)\frac{\partial}{\partial\gamma_{n}}\mu_{n}(\gamma_{n},t)
\end{equation}
The linear flux-force relation which follows is
\begin{equation}
j_{n}(\gamma_{n},t)=-\frac{1}{T}l_{n}(\gamma_{n},t)\frac{\partial}{\partial\gamma_{n}}\mu_{n}(\gamma_{n},t)
\end{equation}
where $l_{n}(\gamma_{n})$ is an Onsager coefficient defined along the reaction coordinate.
 
When the energy barriers are high enough, the system achieves a quasi-stationary state characterized by a uniform current $j_{n}(t)$ given by
\begin{equation}
j_{n}(\gamma_{n},t)=j_{n}(t)
\end{equation}  
If this expression is substituted into Eq. (26), one obtains
\begin{equation}
\dot{S}(t)=-\frac{1}{T}\sum_{n=1}^{N}j_{n}(t)\left[\mu_{n+1}(t)-\mu_{n}(t)\right]
\end{equation}
In the derivation of this equation we used the conditions $\rho_{n}(1,t)=\rho_{n+1}(t)=\rho_{n+1}(0,t)$ which imply that $\mu_{n}(1,t)=\mu_{n+1}(t)=\mu_{n+1}(0,t)$. Equation (30) coincides with Eq. (16) obtained from the discrete analysis.

We now show that the continuous analysis leads to the non-linear law for the observed unfolding rate. For this purpose, we take the fact that the Onsager coefficient $l_{n}(\gamma_{n},t)$ interpreted as a conductivity is proportional to $\rho_{n}(\gamma_{n},t)$. The unfolding rate can be written as
\begin{equation}
j_{n}(t)=-k_{B}\frac{l_{n}(\gamma_{n},t)}{\rho_{n}(\gamma_{n},t)}\rho_{n,st}(\gamma_{n})\frac{\partial}{\partial\gamma_{n}}e^{\frac{\mu_{n}(\gamma_{n},t)}{k_{B}T}}
\end{equation}
We use that $l_{n}(\gamma_{n},t)/\rho_{n}(\gamma_{n},t)\equiv d_{n}$ is independent of the $\gamma_{n}$ coordinate. Furthermore we consider the quasi-stationary limit given in Eq. (27). Dividing this expression by the stationary probability and integrating over the coordinate then gives
\begin{equation}
j_{n}(t)=-D_{n}\left(exp({\frac{\mu_{n+1}(t)}{k_{B}T}})-\exp({\frac{\mu_{n}(t)}{k_{B}T}})\right)
\end{equation}
where
\begin{equation}
D_{n}=\frac{k_{B}d_{n}}{\int_{0}^{1} d\gamma_{n}\rho_{n,st}^{-1}(\gamma_{n})}
\end{equation}
Through the definition of the chemical potential (9), Eq. (30) can be written in the following form
\begin{equation}
j_{n}(t)=-\frac{1}{\tau}\left(\rho_{n+1}(t)\exp({-\frac{g_{s}(n,F)}{k_{B}T}})-\rho_{n}(t)\exp({-\frac{g_{o}(n)}{k_{B}T}})\right)
\end{equation}
where 
\begin{equation}
\frac{1}{\tau}=\frac{D_{n}}{\rho_{n,st}}\exp({\frac{g_{o}(n)}{k_{B}T}})
\end{equation}
In this equation we did not give $\tau$ a subscript $n$. The reason is that ${D_{n}}/{\rho_{n,st}}$ as well as $g_{o}(n)$ are in good approximation independent of $n$. 
By substituting the obtained form (33) for the unfolding rate and the coresponding one for $j_{n+1}(t)$ into the conservation law (3) we obtain the master equation (2). We have then shown that a straightforward analysis in the context of mesoscopic non-equilibrium thermodynamics leads to the observed kinetics of the unfolding process. The theory developed provides also a motivation for why $\tau$ does not depend on $n$.

\section{Discussion and Conclusions}

In this paper, we have shown that a pertinent non-equilibrium thermodynamics treatment of the RNA unfolding yields the same master equation as the one proposed in a kinetic treatment of the problem. We summarize the reasons why our method can give a general description of kinetic processes taking place in single-molecules. 

We view each step in the unfolding of the molecule, not as a sudden switch, but as occuring via many small intermediate steps leading the system through a virtual continuum of states. When a non-equilibrium thermodynamics scheme is applied, not to the global transformation but to these small switches, the resulting linear contributions to the rate integrate to give the observed non-linear behavior. A thermodynamic description of the kinetic process is legitimate when the intermediate states persist on the time scale considered. Therefore the states along the $\gamma$-coordinates can be considered as thermodynamic states. This condition is fulfilled when the system has time enough to equilibrate locally during the transformation, and the activated jump can be viewed as a diffusion process. Following these ideas, embedded in the theory of mesoscopic non-equilibrium thermodynamics, we have analyzed the RNA unfolding under tension, arriving at the result that kinetic and mesoscopic non-equilibrium thermodynmics descriptions are equivalent. 
 
Classical non-equilibrium thermodynamics, however, performs an analysis of these processes in terms of only two states of the molecule: the initial state and the final state reached after the transformation has taken place. Since the laws derived within this framework are linear, it necessarily leads to linear relationships for the rates as a function of chemical potential differences. These relations constitute only aproximations to the kinetic laws observed \cite{Hill2} and therefore can only provide a partial description of the process.

\subsubsection*{Acknowledgment}

The authors are grateful for the Storforsk grant no 167336/V30 from the
Norwegian Research Council.

\pagebreak


\begin{thebibliography}{99}
\bibitem{Tinoco.Bustamante}Tinoco, I.; Bustamante, C. J. Molec. Bio. \textbf{1999}, 
293, 271.
\bibitem{Cocco} Cocco, S.; Monasson, R.; Marko, J. Euro. Phys. J. E \textbf{2003},
10, 153.
\bibitem{Vieregg.Tinoco} Tinoco, I.; Vieregg, J.R. Molecular Physics \textbf{2006},
104, 1343.
\bibitem{Tinoco} Tinoco, I. Annu. Rev. Biophys. Biomol. Struct. \textbf{2004},
33, 363.
\bibitem{looping.current} Vilar, J.M.G.; Saiz, L. Curr. Opin. Genet. Dev. \textbf{2005},
15, 136.
\bibitem{looping.pnas}Saiz, L.; Rubi, J.M.; Vilar, J.M.G. Proc. Natl. Acad. Sci \textbf{2005},
102, 17642.
\bibitem{Hill} Hill, T.L. Thermodynamics of small systems; Dover: New York,
1994.
\bibitem{Callen} Callen, H.B. Thermodynamics and an Introduction to
Thermostatistics; John Wiley and Sons, Inc.: New York, 1985.
\bibitem{stretching} Rubi, J.M.; Bedeaux, D.; Kjelstrup, S. J. Phys. Chem. B  \textbf{2006}, 110, 12733.
\bibitem{Onsager.Machlup} Onsager, L.; Machlup, S. Phys. Rev. \textbf{1953},
91, 1505.
\bibitem{Caplan}  Caplan, S. R. and Essig. A. Bioenergetics and linear
nonequilibrium thermodynamics. The steady state; Harvard University Press:
Cambridge, MA, 1983.
\bibitem{Westerhoff}  Westerhoff, H.V. and van Dam, K. Thermodynamics
and control of biological free-energy transduction; Elsevier: Amsterdam, 1987.
\bibitem{dGM} de Groot, S.R.; Mazur, P. Non-Equilibrium thermodynamics,
Dover: New York, 1984.
\bibitem{Hill2}Hill, T.L. Free energy transduction and biochemical cycle kinetics,
Springer Verlag: New York, 1989.
\bibitem{Qian} Qian, H.  J. Phys. Chem. B \textbf{2006}, 110, 15063.
\bibitem{mnet} Reguera, D.; Rubi, J.M.; Vilar, J.M.G. J. Phys. Chem. B \textbf{2005},
109, 21502.
\bibitem{JTB} Kjelstrup, S.; Rubi, J.M.; Bedeaux, D.  J. Theor. Biol. \textbf{2005}, 234, 7.
\bibitem{PCCP} Kjelstrup, S.; Rubi, J.M.; Bedeaux, D. Phys. Chem. Chem. Phys. \textbf{2005}, 7 4009.
\bibitem{Doi.Edwards} Doi, M.; Edwards, S. F. The theory of polymer dynamics;
Clarendon: Oxford, 1986.
\bibitem{master.nucleation} Reguera, D.; Rubi, J.M. Physica A \textbf{1998},
259,10.


\end{thebibliography}
\end{document}